\documentclass[preprint,showpacs,preprintnumbers,amsmath,amssymb]{revtex4}

\newcommand\rf[1]{(\ref{eq:#1})}
\newcommand\lab[1]{\label{eq:#1}}
\newcommand\nonu{\nonumber}
\newcommand\br{\begin{eqnarray}}
\newcommand\er{\end{eqnarray}}
\newcommand\be{\begin{equation}}
\newcommand\ee{\end{equation}}

\newcommand\lb{\lbrack}
\newcommand\rb{\rbrack}

\newcommand\llb{\left\lbrack}
\newcommand\rrb{\right\rbrack}

\newcommand\lcurl{\left\{}
\newcommand\rcurl{\right\}}
\renewcommand\({\left(}
\renewcommand\){\right)}
\newcommand\bv{\bigm\vert}               
\newcommand\bgv{\bigg\vert}              

\newcommand\bc{\begin{center}}
\newcommand\ec{\end{center}}

\newcommand\partder[2]{\frac{{\partial {#1}}}{{\partial {#2}}}}

\renewcommand\a{\alpha}
\renewcommand\b{\beta}

\newcommand\vareps{\varepsilon}
\newcommand\g{\gamma}
\newcommand\G{\Gamma}

\newcommand\h{\frac{1}{2}}
\renewcommand\k{\kappa}
\renewcommand\l{\lambda}

\newcommand\m{\mu}
\newcommand\n{\nu}

\newcommand\vp{\varphi}
\renewcommand\P{\Phi}
\newcommand\pa{\partial}

\newcommand\pr{\prime}

\newcommand\s{\sigma}

\renewcommand\t{\tau}
\renewcommand\th{\theta}

\newcommand{\ct}[1]{\cite{#1}}
\newcommand{\bib}[1]{\bibitem{#1}}
\newcommand\PRL[3]{\textsl{Phys. Rev. Lett.} \textbf{#1}, #3 (#2)}
\newcommand\NPB[3]{\textsl{Nucl. Phys.} \textbf{B#1}, #3 (#2)}

\newcommand\CMP[3]{\textsl{Commun. Math. Phys.} \textbf{#1}, #3 (#2)}
\newcommand\PRD[3]{\textsl{Phys. Rev.} \textbf{D#1}, #3 (#2)}

\newcommand\PLB[3]{\textsl{Phys. Lett.} \textbf{#1B}, #3 (#2)}
\newcommand\CQG[3]{\textsl{Class. Quantum Grav.} \textbf{#1}, #3 (#2)}

\newcommand\PTP[3]{\textsl{Prog. Theor. Phys.} \textbf{#1}, #3 (#2)}

\newcommand\IJMPA[3]{\textsl{Int. J. Mod. Phys.} \textbf{A#1}, #3 (#2)}

\newcommand\ydot{\stackrel{.}{y}}

\begin{document}


\title{``Mass Inflation'' With Lightlike Branes}

\author{Eduardo I. Guendelman and Alexander B. Kaganovich}%
\email{guendel@bgumail.bgu.ac.il , alexk@bgumail.bgu.ac.il}
\affiliation{%
Department of Physics, Ben-Gurion University of the Negev \\
P.O.Box 653, IL-84105 ~Beer-Sheva, Israel
}%

\author{Emil R. Nissimov and Svetlana J. Pacheva}%
\email{nissimov@inrne.bas.bg , svetlana@inrne.bas.bg}
\affiliation{%
Institute for Nuclear Research and Nuclear Energy,
Bulgarian Academy of Sciences \\
Boul. Tsarigradsko Chausee 72, BG-1784 ~Sofia, Bulgaria
}%

\begin{abstract}
We discuss properties of a new class of $p$-brane models, describing
intrinsically {\em lightlike} branes for any world-volume dimension, in various
gravitational backgrounds of interest in the context of black hole physics.
One of the characteristic features of these lightlike $p$-branes is that the
brane tension appears as an additional nontrivial dynamical world-volume degree 
of freedom. Codimension one lightlike brane dynamics requires that bulk space 
with a bulk metric of spherically symmetric type must possess
an event horizon which is automatically occupied by the lightlike
brane while its tension evolves exponentially with time. The latter
phenomenon is an analog of the well known ``mass inflation'' effect in
black holes.

\vspace{.2in} \noindent
Keywords: {\em non-Nambu-Goto lightlike $p$-branes; dynamical brane tension;
black hole's horizon ``straddling''; exponential inflation/deflation}
\end{abstract}

\pacs{11.25.-w, 04.70.-s, 04.50.+h}

\maketitle

\section{Introduction}

The behavior of matter near horizons of black holes has been the subject of
several interesting investigations \ct{mat-hor-1}-\ct{poisson-israel-2}. One
particularly intriguing effect was the ``mass inflation''
\ct{poisson-israel-1}-\ct{poisson-israel-2} which takes place, for example, 
for matter accumulating (blue shifting) near the inner Reissner-Nordstr{\"o}m horizon.

In the context of the problem where we consider matter living close to, or in
fact on, the horizons of black holes, the notion of lightlike branes becomes
particularly relevant. Let us recall that
lightlike branes (\textsl{LL-branes}, for short) are of particular interest in 
general relativity primarily due to their role: (i) in describing impulsive lightlike 
signals arising in cataclysmic astrophysical events \ct{barrabes-hogan};
(ii) as basic ingredients in the so called ``membrane paradigm'' 
theory \ct{membrane-paradigm} of black hole physics; (iii) in the
context of the thin-wall description of domain walls coupled to 
gravity \ct{Israel-66-1}-\ct{Barrabes-Israel-Hooft-2}.

More recently, \textsl{LL-branes} became significant also in the context of
modern non-perturbative string theory, in particular, as the so called
$H$-branes describing quantum horizons (black hole and cosmological)
\ct{kogan-01}, as well appearing as Penrose limits of baryonic $D(=$Dirichlet)
branes \ct{mateos-02}.

In the original papers \ct{Israel-66-1}-\ct{Barrabes-Israel-Hooft-2} \textsl{LL-branes}
in the context of gravity and cosmology have been extensively studied from a 
phenomenological point of view, \textsl{i.e.}, by introducing them without specifying
the Lagrangian dynamics from which they may originate.
In a recent paper \ct{barrabes-israel-05} brane actions in terms of their 
pertinent extrinsic geometry have been proposed which generically describe 
non-lightlike branes, whereas the lightlike branes are treated as a limiting case.
On the other hand, we have proposed in a series of recent papers 
\ct{will-brane-1}-\ct{will-brane-7} a new class of concise Lagrangian
actions, among them -- {\em Weyl-conformally invariant} ones, providing a
derivation from first principles of the \textsl{LL-brane} dynamics.

In Section 2 of the present paper we extend our previous
construction (which was restricted to odd world-volume dimensions) to the case of 
\textsl{LL-brane} actions for {\em arbitrary} world-volume dimensions. 

In Section 3 we discuss the properties of \textsl{LL-brane} 
dynamics 
moving as test brane in generic gravitational backgrounds. 
The case with two extra 
dimensions (codimension two \textsl{LL-branes}) was studied in a recent paper
\ct{varna-07} from the point of view of ``braneworld'' scenarios
\ct{brane-world-1}-\ct{brane-world-6} (for a review, see 
\ct{brane-world-rev-1}-\ct{brane-world-rev-2}). 
Unlike conventional braneworlds, where the underlying branes are of Nambu-Goto type 
(\textsl{i.e.}, describing massive brane modes) and in their ground state they 
position themselves at some
fixed point in the extra dimensions of the bulk space-time, our lightlike
braneworlds perform in the ground state non-trivial motions in the extra
dimensions -- planar circular, spiral winding \textsl{etc} depending on the
topology of the extra dimensions. In the present paper we concentrate on
the special case of codimension one \textsl{LL-branes} which is qualitatively
different and needs separate study.
Here the consistency of \textsl{LL-brane} dynamics as test brane moving in
external gravitational fields dictates that the bulk space-time 
with a bulk metric of spherically symmetric type (see Eq.\rf{spherical-metric} below)
must possess an event horizon which is automatically occupied by the \textsl{LL-brane}
(``horizon straddling'' according to the terminology of 
ref.\ct{Barrabes-Israel-Hooft-1}).
This is a generalization for any $(p+1)$ world-volume dimensions of the results 
previously obtained in refs.\ct{will-brane-1}-\ct{will-brane-7} for lightlike
membranes ($p=2$) in $D=4$ bulk space-time.

In Section 4 we study several cases of ``horizon straddling'' solutions
obtained from our \textsl{LL-brane} world-volume action \rf{LL-brane}. For
the inner Reissner-Nordstr{\"o}m horizon we find a time symmetric ``mass
inflation'' scenario, which also holds for de Sitter horizon. In this case
the dynamical tension of the \textsl{LL-brane} blows up as time approaches
$\pm \infty$ due to its exponential quadratic time dependence.
For the Schwarzschild and the outer Reissner-Nordstr{\"o}m horizons, on the
other hand, we obtain ``mass deflationary'' scenarios where the dynamical
\textsl{LL-brane} tension vanishes at large positive or large negative times.
Another set of solutions with asymmetric (w.r.t. $t \to -t$) exponential
linear time dependence of the \textsl{LL-brane} tension also exists. 
In the latter case, by fine tuning one can obtain a constant
time-independent brane tension as a special case.

\section{World-Volume Actions of Lightlike Branes}

We propose the following reparametrization invariant action describing
intrinsically lightlike $p$-branes for any world-volume dimension $(p+1)$
(for previous versions, cf.\ct{will-brane-1}-\ct{will-brane-7}):
\be
S = - \int d^{p+1}\s \,\P (\vp)
\Bigl\lb \h \g^{ab} \pa_a X^{\m} \pa_b X^{\n} G_{\m\n}(X) - L\!\( F^2\)\Bigr\rb
\lab{LL-brane}
\ee
using notions and notations as follows:

\begin{itemize}
\item
Alternative non-Riemannian integration measure density $\P (\vp)$ (volume form) on
the $p$-brane world-volume manifold:
\be
\P (\vp) \equiv \frac{1}{(p+1)!} \vareps_{I_1\ldots I_{p+1}}
\vareps^{a_1\ldots a_{p+1}} \pa_{a_1} \vp^{I_1}\ldots \pa_{a_{p+1}} \vp^{I_{p+1}}
\lab{mod-measure-p}
\ee
instead of the usual $\sqrt{-\g}$. Here $\lcurl \vp^I \rcurl_{I=1}^{p+1}$ are
auxiliary world-volume scalar fields; $\g_{ab}$ ($a,b=0,1,{\ldots},p$)
denotes the intrinsic Riemannian metric on the world-volume, and
$\g = \det\Vert\g_{ab}\Vert$.
Note that $\g_{ab}$ is {\em independent} of $\vp^I$.
\item
$X^\m (\s)$ are the $p$-brane embedding coordinates in the bulk
$D$-dimensional space time with bulk Riemannian metric
$G_{\m\n}(X)$; $\m,\n = 0,1,\ldots ,D-1$, 
$(\s)\equiv \(\s^0 \equiv \t,\s^1,\ldots ,\s^p\)$, $\pa_a \equiv \partder{}{\s^a}$.
\item
Auxiliary $(p-1)$-rank antisymmetric tensor gauge field $A_{a_1\ldots a_{p-1}}$
on the world-volume with $p$-rank field-strength and its dual:
\be
F_{a_1 \ldots a_{p}} = p \pa_{[a_1} A_{a_2\ldots a_{p}]} \quad ,\quad
F^{\ast a} = \frac{1}{p!} \frac{\vareps^{a a_1\ldots a_p}}{\sqrt{-\g}}
F_{a_1 \ldots a_{p}}  \; .
\lab{p-rank}
\ee
\item
$L\!\( F^2\)$ is {\em arbitrary} function of $F^2$ with the short-hand notation:
\be
F^2 \equiv F_{a_1 \ldots a_{p}} F_{b_1 \ldots b_{p}} 
\g^{a_1 b_1} \ldots \g^{a_p b_p} \; .
\lab{F2-id}
\ee
\end{itemize}

Let us note the simple identity:
\be
F_{a_1 \ldots a_{p-1}b}F^{\ast b} = 0 \; ,
\lab{F-id}
\ee
which will play a crucial role in the sequel.

\textbf{Remark 1.} For the special choice
$L\!\( F^2\)= \( F^2\)^{1/p}$ the action \rf{LL-brane} becomes 
manifestly invariant under {\em Weyl (conformal) symmetry}: 
$\g_{ab}\!\! \longrightarrow\!\! \g^{\pr}_{ab} = \rho\,\g_{ab}$,
$\vp^{I} \longrightarrow \vp^{\pr\, I} = \vp^{\pr\, I} (\vp)$ with Jacobian 
$\det \Bigl\Vert \frac{\pa\vp^{\pr\, I}}{\pa\vp^J} \Bigr\Vert = \rho$.
In what follows we will consider the generic Weyl {\em non}-invariant case.

\textbf{Remark 2.} In our previous papers \ct{will-brane-1}-\ct{will-brane-7}
we have used a different form for the Lagrangian of the auxiliary world-volume 
gauge field in the brane action \rf{LL-brane}: 
\be
L(F^2) = \sqrt{F_{ab} F_{cd} \g^{ac} \g^{bd}} \;\;\; \mathrm{with} \;\;
F_{ab} = \pa_a A_b - \pa_b A_a \; ,
\lab{A-odd-dim}
\ee
\textsl{i.e.}, with  ordinary vector gauge field for any $p$. However, it has been
shown in ref.\ct{will-brane-7} that for the choice \rf{A-odd-dim}
the action \rf{LL-brane} describes consistent brane dynamics only for {\em odd} 
$(p+1)$ world-volume dimensions. This is due to the following relation 
(Eq.(13) in ref.\ct{will-brane-7}, which is a consequences from
the equation of motion w.r.t. $\g_{ab}$ -- the counterpart of Eq.\rf{gamma-eqs}
below):
\be
\det\Vert \(\pa_a X \pa_b X\) \Vert = \( -4 L^\pr\! (F^2)\)^{p+1}
\(-\det\Vert\g_{ab}\Vert\) \(\det\Vert iF_{ab}\Vert\)^2 \; .
\lab{det-gamma-eqs}
\ee
The latter relation implies that for $(p+1)=$even world-volume dimensions the 
r.h.s. of \rf{det-gamma-eqs} is strictly positive (because of the Lorentzian signature
of the intrinsic metric $\g_{ab}$) contradicting the requirement that
the determinant of the induced metric in the l.h.s of \rf{det-gamma-eqs}
should be negative conforming with the Lorentzian signatures of both world-volume
and embedding space-time metrics. Henceforth, we will employ our new action 
\rf{LL-brane} with the $(p-1)$-rank auxiliary world-volume antisymmetric tensor 
gauge fields \rf{p-rank}.

Rewriting the action \rf{LL-brane} in the following equivalent form:
\be
S = - \int d^{p+1}\!\s \,\chi \sqrt{-\g}
\Bigl\lb \h \g^{ab} \pa_a X^{\m} \pa_b X^{\n} G_{\m\n}(X) - L\!\( F^2\)\Bigr\rb
\quad, \quad
\chi \equiv \frac{\P (\vp)}{\sqrt{-\g}}
\lab{LL-brane-chi}
\ee
with $\P (\vp)$ the same as in \rf{mod-measure-p},
we find that the composite field $\chi$ plays the role of a {\em dynamical
(variable) brane tension}. Let us note that the notion of dynamical brane 
tension has previously appeared in different contexts in 
refs.\ct{townsend-etal-1}-\ct{townsend-etal-3}.

Before proceeding, let us mention that both the auxiliary world-volume scalars 
$\vp^I$ entering the non-Riemannian integration measure density
\rf{mod-measure-p}, as well as the intrinsic world-volume metric $\g_{ab}$ are
{\em non-dynamical} degrees of freedom in the action \rf{LL-brane},
or equivalently, in \rf{LL-brane-chi}. Indeed, there are no (time-)derivates
w.r.t. $\g_{ab}$, whereas the action \rf{LL-brane} (or \rf{LL-brane-chi}) is
{\em linear} w.r.t. the velocities $\stackrel{.}{\vp}^I$. Thus,
\rf{LL-brane} (or \rf{LL-brane-chi}) is a constrained dynamical system,
\textsl{i.e.}, a system with gauge symmetries including the gauge symmetry 
under world-volume reparametrizations (about the Hamiltonian treatment of
\rf{LL-brane}, see the remarks after Eq.\rf{gamma-eqs} below). 
On the other hand, 
the dynamical brane tension $\chi$ \rf{LL-brane-chi},
being a ratio of two world-volume scalar densities, is itself a well-defined
reparametrization-covariant {\em world-volume scalar field}.

Introducing a short-hand notation for the induced metric:
\be
\(\pa_a X \pa_b X\) \equiv \pa_a X^\m \pa_b X^\n G_{\m\n}  \; .
\lab{metric-short-hand}
\ee
we can write the equations of motion obtained from \rf{LL-brane} w.r.t. 
measure-building auxiliary scalars $\vp^I$ and $\g^{ab}$ as:
\be
\h \g^{cd}\(\pa_c X \pa_d X\) - L\!\( F^2\) = M \; ,
\lab{phi-eqs}
\ee
where $M$ is an integration constant;
\be
\h\(\pa_a X \pa_b X\) - p L^\pr\!\( F^2\) F_{a a_1\ldots a_{p-1}}
\g^{a_1 b_1}\ldots \g^{a_{p-1} b_{p-1}} F_{b b_1 \ldots b_{p-1}} = 0 \; .
\lab{gamma-eqs}
\ee
Since, as mentioned above, both $\vp^I$ and $\g^{ab}$ are non-dynamical
degrees of freedom, both Eqs.\rf{phi-eqs}--\rf{gamma-eqs} are in fact
{\em non-dynamical constraint} equations (no second-order time derivatives
present). Their meaning as constraint
equations is best understood within the framework of the Hamiltonian formalism 
for the action \rf{LL-brane} (or \rf{LL-brane-chi}). The latter can be developed
in strict analogy with the Hamiltonian formalism for a simpler class of
modified $p$-brane models based on the alternative non-Riemannian
integration measure density \rf{mod-measure-p}, which was previously proposed 
in \ct{m-string} (for details, we refer to Sections 2 and 3 of \ct{m-string}).
In particular, Eqs.\rf{gamma-eqs} can be viewed as $p$-brane 
analogues of the string Virasoro constraints.

Thus, Eqs.\rf{phi-eqs}--\rf{gamma-eqs} are particular manifestation
in the case of \rf{LL-brane} of the general property in any dynamical system
with gauge symmetries, \textsl{i.e.}, a system with constraints a'la Dirac
\ct{gauge-sys-1}-\ct{gauge-sys-3} -- variation of the action w.r.t. non-dynamical 
degrees of freedom (Lagrange multipliers) yields non-dynamical constraint equations.

Taking the trace in \rf{gamma-eqs} and comparing with \rf{phi-eqs} 
implies the following crucial relation for the Lagrangian function $L\( F^2\)$: 
\be
L\!\( F^2\) - p F^2 L^\pr\!\( F^2\) + M = 0 \; ,
\lab{L-eq}
\ee
which determines $F^2$ \rf{F2-id} on-shell as certain function of the integration
constant $M$ \rf{phi-eqs}, \textsl{i.e.}
\be
F^2 = F^2 (M) = \mathrm{const} \; .
\lab{F2-const}
\ee

The second and most profound consequence of Eqs.\rf{gamma-eqs} is due to
the identity \rf{F-id} which implies that the induced metric 
\rf{metric-short-hand} on the world-volume of the $p$-brane model \rf{LL-brane} 
is {\em singular} (as opposed to the induced metric in the case of 
ordinary Nambu-Goto branes):
\br
\(\pa_a X \pa_b X\) F^{\ast b} = 0 \quad ,\quad \mathrm{i.e.}\;\;
\(\pa_F X \pa_F X\) = 0 \;\; ,\;\; \(\pa_{\perp} X \pa_F X\) = 0
\lab{LL-constraints}
\er
where $\pa_F \equiv F^{\ast a} \pa_a$ and $\pa_{\perp}$ are derivatives 
along the tangent vectors in the complement of $F^{\ast a}$.

Thus, we arrive at the following important conclusion: every point on the 
surface of the $p$-brane \rf{LL-brane} moves with the speed of light
in a time-evolution along the vector-field $F^{\ast a}$ which justifies the
name {\em LL-brane} (Lightlike-brane) model for \rf{LL-brane}.

Before proceeding let us point out that we can add \ct{varna-07} to the 
\textsl{LL-brane} action \rf{LL-brane} natural couplings to bulk Maxwell and 
Kalb-Ramond gauge fields. The latter do not affect Eqs.\rf{phi-eqs} and 
\rf{gamma-eqs}, so that the conclusions about on-shell
constancy of $F^2$ \rf{F2-const} and the lightlike nature \rf{LL-constraints} of the 
$p$-branes under consideration remain unchanged.

The remaining equations of motion w.r.t. auxiliary
world-volume gauge field $A_{a_1 \ldots a_{p-1}}$ and $X^\m$ produced by the
action \rf{LL-brane} read:
\be
\pa_{[a}\( F^{\ast c} \g_{b]c}\, \chi L^\pr(F^2)\) = 0 \; ;
\lab{A-eqs}
\ee
\br
\pa_a \(\chi \sqrt{-\g} \g^{ab} \pa_b X^\m\) + 
\chi \sqrt{-\g} \g^{ab} \pa_a X^\n \pa_b X^\l \G^\m_{\n\l}(X) = 0
\lab{X-eqs}
\er
Here $\chi$ is the dynamical brane tension as in \rf{LL-brane-chi},
\be
\G^\m_{\n\l}=\h G^{\m\k}\(\pa_\n G_{\k\l}+\pa_\l G_{\k\n}-\pa_\k G_{\n\l}\)
\lab{affine-conn}
\ee
is the Christoffel connection for the external metric,
and $L^\pr(F^2)$ denotes derivative of $L(F^2)$ w.r.t. the argument $F^2$.


Invariance under world-volume reparametrizations allows to introduce the
standard synchronous gauge-fixing conditions:
\be
\g^{0i} = 0 \;\; (i=1,\ldots,p) \quad ,\quad \g^{00} = -1 \; .
\lab{gauge-fix}
\ee
Also, in what follows we will use a natural ansatz for the auxiliary world-volume
gauge field-strength:
\be
F^{\ast i}= 0 \;\; (i=1,{\ldots},p) \quad ,\quad \mathrm{i.e.} \;\;
F_{0 i_1 \ldots i_{p-1}} = 0 \; ,
\lab{F-ansatz}
\ee
the only non-zero component of the dual field-strength being:
\br
F^{\ast 0}= \frac{1}{p!} \frac{\vareps^{i_1 \ldots i_p}}{\sqrt{\g^{(p)}}}
F_{i_1 \ldots i_p} \; ,
\lab{F-ansatz-1} \\
\g^{(p)} \equiv \det\Vert\g_{ij}\Vert\;\; (i,j=1,\ldots,p) \quad ,\quad
F^2 = p! \( F^{\ast 0}\)^2 = \mathrm{const}  \; .
\nonu
\er
According to \rf{LL-constraints} the meaning of the ansatz \rf{F-ansatz} is
that the lightlike direction $F^{\ast a} \pa_a \simeq \pa_0 \equiv \pa_\t$,
\textsl{i.e.}, it coincides with the brane proper-time direction.
Biancchi identity $\nabla_a F^{\ast a}=0$ together with 
\rf{F-ansatz}--\rf{F-ansatz-1} implies:
\be
\pa_0 F_{i_1 \ldots i_p} = 0 \;\; \longrightarrow \;\; 
\pa_0 \sqrt{\g^{(p)}} = 0  \; .
\lab{gamma-p-const}
\ee

Using \rf{gauge-fix} and \rf{F-ansatz} the equations of motion \rf{gamma-eqs},
\rf{A-eqs} and \rf{X-eqs} acquire the form, respectively:
\be
\(\pa_0 X\,\pa_0 X\) = 0 \quad ,\quad \(\pa_0 X\,\pa_i X\) = 0 \quad ,\quad
\(\pa_i X\,\pa_j X\) - 2a_0\, \g_{ij} = 0
\lab{gamma-eqs-0}
\ee
(Virasoro-like constraints), where the $M$-dependent constant $a_0$: 
\be
a_0 \equiv F^2 L^\pr (F^2)\bv_{F^2 = F^2(M)}
\lab{a0-const}
\ee
must be strictly positive;
\be
\pa_i \chi = 0 \; ;
\lab{A-eqs-0}
\ee
\br
-\sqrt{\g^{(p)}} \pa_0 \(\chi \pa_0 X^\m\) +
\pa_i\(\chi\sqrt{\g^{(p)}} \g^{ij} \pa_j X^\m\)
\nonu \\
+ \chi\sqrt{\g^{(p)}} \(-\pa_0 X^\n \pa_0 X^\l + \g^{kl} \pa_k X^\n \pa_l X^\l\)
\G^\m_{\n\l} = 0 \; .
\lab{X-eqs-0}
\er

\section{Lightlike Brane Dynamics in Gravitational Backgrounds}

Let us split the bulk space-time coordinates as:
\br
\( X^\m\) = \( x^a, y^\a\) \equiv \( x^0 \equiv t, x^i, y^\a\)
\lab{coord-split} \\
a=0,1,\ldots, p \;\; ,\;\; i=1,\ldots, p \;\; ,\;\;
\a = 1,\ldots, D-(p+1)
\nonu
\er
and consider background metrics $G_{\m\n}$ of the form:
\be
ds^2 = - A(t,y)(dt)^2 + C(t,y) g_{ij}(\vec{x}) dx^i dx^j + 
B_{\a\b}(t,y) dy^\a dy^\b \; .
\lab{nonstatic-metric}
\ee

Here we will discuss the simplest non-trivial ansatz for the \textsl{LL-brane}
embedding coordinates:
\be
X^a \equiv x^a = \s^a \quad, \quad X^{p+\a} \equiv y^\a = y^\a (\t) \quad,
\quad \t \equiv \s^0  \; .
\lab{X-ansatz}
\ee
With \rf{nonstatic-metric} and \rf{X-ansatz}, the constraint Eqs.\rf{gamma-eqs-0} 
yield:
\be
- A + B_{\a\b} \ydot^\a \ydot^\b = 0 \quad ,\quad
C g_{ij} - 2 a_0 \g_{ij} = 0 \; ,
\lab{gamma-eqs-1}
\ee
where $\ydot^\a \equiv \frac{d}{d\t} y^\a$. Second Eq.\rf{gamma-eqs-1}
together with the last relation in \rf{gamma-p-const} implies:
\be
\frac{d}{d\t} C(y(\t)) = 
\Bigl(\pa_t C + \ydot^\a\!\!\partder{C}{y^\a}\Bigr)\bgv_{t=\t,y=y(\t)}=0 \; .
\lab{C-rel}
\ee

The second-order Eqs.\rf{X-eqs-0} for $X^0 \equiv t$ and $X^{p+\a} \equiv y^\a$
yield accordingly:
\br
\pa_\t \chi + \frac{\chi}{A}\Bigl\lb \h\pa_t A + \ydot^\a\!\!\partder{A}{y^\a}
+ \h \ydot^\a \ydot^\b \pa_t B_{\a\b} + 
\frac{p\, a_0}{C}\,\ydot^\a\!\!\partder{C}{y^\a} \Bigr\rb\bgv_{t=\t,y=y(\t)} = 0
\; ,
\lab{X0-eq} 
\er
\br
\pa_\t \(\chi \ydot^\a\) + 
\chi \Bigl\lb B^{\a\b} \Bigl(\h \partder{A}{y^\a} + \ydot^\g \pa_t B_{\b\g} + 
\frac{p\, a_0}{C}\,\partder{C}{y^\b}\Bigr) + 
\ydot^\b \ydot^\g \G^\a_{\b\g}\Bigr\rb\bgv_{t=\t,y=y(\t)} = 0  \; ,
\lab{y-eqs}
\er
where $\G^\a_{\b\g}$ is the Christoffel connection for the metric $B_{\a\b}$
in the extra dimensions (cf. \rf{nonstatic-metric}).

\textsl{LL-brane} equations \rf{gamma-eqs-1}--\rf{y-eqs} for codimension two
(\textsl{i.e.}, for $D-(p+1)\!=\!2$) have been studied in ref.\ct{varna-07} from
the braneworld point of view. The case of codimension one \textsl{LL-branes}
moving in gravitational backgrounds (\textsl{i.e.}, for $D=p+2$) is
qualitatively different and is the subject of the discussion in what follows.

In the latter case the metric \rf{nonstatic-metric} acquires the form of a general 
spherically symmetric metric:
\be
ds^2 = - A(t,y)(dt)^2 + C(t,y) g_{ij}(\vec{\th}) d\th^i d\th^j + B (t,y) (dy)^2
\; ,
\lab{spherical-metric}
\ee
where $\vec{x} \equiv \vec{\th}$ are the angular coordinates
parametrizing the sphere $S^p$. 

Eqs.\rf{gamma-eqs-1}--\rf{X0-eq} now take the form: 
\br
-A + B \ydot^2 = 0 \;\; ,\; \mathrm{i.e.}\;\; \ydot = \pm \sqrt{\frac{A}{B}}
\quad ,\quad
\pa_t C + \ydot \pa_y C = 0
\lab{r-const} \\
\pa_\t \chi + \chi \Bigl\lb \pa_t \ln \sqrt{AB} 
\pm \frac{1}{\sqrt{AB}} \Bigl(\pa_y A + p\, a_0 \pa_y \ln C\Bigr)\Bigr\rb = 0
\; ,
\lab{X0-eq-1}
\er
whereas Eq.\rf{y-eqs} becomes a consequence of the above ones.

In what follows we will consider the following subclasses of background
metrics \rf{spherical-metric}:

(i) Static spherically symmetric metrics in standard coordinates:
\be
A = A(y) \;\; ,\;\; B(y) = A^{-1}(y)\;\; ,\;\; C(y) = y^2 \; ,
\lab{standard-spherical}
\ee
where $y\equiv r$ is the radial-like coordinate. In the case of 
\rf{standard-spherical}, Eqs.\rf{r-const} imply:
\be
\ydot = 0 \;\; ,\;\; \mathrm{i.e.}\;\; y(\t) = y_0 = \mathrm{const} \quad, \quad
A(y_0) = 0 \; .
\lab{horizon-standard}
\ee
In other words, the equations of motion of the \textsl{LL-brane} require that 
the latter positions itself on a spherical-like hypersurface 
(second Eq.\rf{horizon-standard}) in the bulk space-time which in addition
must be a horizon of the background metric (last Eq.\rf{horizon-standard},
cf. \rf{spherical-metric}).

The next Eq.\rf{X0-eq-1} reduces in the case of \rf{standard-spherical} to:
\be
\pa_\t \chi \pm \chi \Bigl(\pa_y A\bv_{y=y_0} + \frac{2 p\, a_0}{y_0}\Bigr)
= 0
\lab{chi-eq-standard}
\ee
with the obvious solution:
\be
\chi (\t) = \chi_0 
\exp\Bigl\{\mp \t \Bigl(\pa_y A\bv_{y=y_0} + \frac{2 p\, a_0}{y_0}\Bigr)\Bigr\}
\quad ,\quad \chi_0 = \mathrm{const} \; .
\lab{chi-eq-standard-sol}
\ee
Thus, we find a time-asymmetric solution for the dynamical brane tension which
(upon appropriate choice of the signs $(\mp)$ depending on the sign of the
constant factor in the exponent on the r.h.s. of \rf{chi-eq-standard-sol})
exponentially ``inflates'' for large times. In the particular case of fine
tuning of parameters:
\be
\pa_y A\bv_{y=y_0} + \frac{2 p\, a_0}{y_0} = 0
\lab{fine-tune}
\ee
we obtain a constant solution $\chi = \chi_0$.

(ii) Spherically symmetric metrics in Kruskal-like coordinates:
\be
A=B \;\; ,\;\; A=A\(y^2 - t^2\) \;\; ,\;\; C=C\(y^2 - t^2\)\; ,
\lab{kruskal-like}
\ee
where $(t,y)$ play the role of Kruskal's $(v,u)$ coordinates for Schwarzschild
metrics \ct{kruskal-1}-\ct{kruskal-2}. In the case of \rf{kruskal-like}, 
Eqs.\rf{r-const} xyield:
\be
\ydot = \pm 1 \;\;, \;\; \mathrm{i.e.}\;\; y(\t) = \pm \t \quad ,\quad
\(y^2 - t^2\)\bv_{t=\t,y=y(\t)} = 0 \; ,
\lab{horizon-kruskal}
\ee
\textsl{i.e.}, again the \textsl{LL-brane} locates itself {\em automatically} on the
horizon. Eq.\rf{X0-eq-1} reduces accordingly to:
\br
\pa_\t \chi + \t \frac{2p\, a_0\, C^{\pr}(0)}{A(0)C(0)}\,\chi = 0 \; ,
\lab{chi-eq-kruskal} \\
\mathrm{i.e.}\;\; \chi (\t) = \chi_0 
\exp\Bigl\{ -\t^2\,\frac{p\, a_0\, C^{\pr}(0)}{A(0)C(0)}\Bigr\}
\lab{chi-eq-kruskal-sol}
\er
Thus, we find a time-symmetric ``inflationary'' or ``deflationary'' solution
for the dynamical brane tension depending on the sign of the constant factor
in the exponent on the r.h.s. of \rf{chi-eq-kruskal-sol}.

(iii) ``Cosmological''-type metrics:
\be
A=1 \;\; ,\;\; B=S^2(t) \;\; ,\;\; C=S^2(t)\, f^2(y)\; ,
\lab{cosmolog-type}
\ee
\textsl{i.e.}:
\be
ds^2 = - (dt)^2 + S^2(t)\llb (dy)^2 + f^2 (y) g_{ij}(\vec{\th}) d\th^i d\th^j\rrb \,
\lab{FRW-metric}
\ee
with $\th^i$ parametrizing the $p$-dimensional sphere $S^p$.
In this case Eqs.\rf{r-const} give:
\be
\ydot = \pm \frac{1}{S(\t)} \quad ,\quad
C\bv_{t=\t,y=y(\t)}\equiv S^2(\t)\, f^2(y(\t)) = \frac{1}{c_0^2} \;\; ,\;\;
c_0 = \textrm{const} \; ,
\lab{horizon-cosmolog}
\ee
implying:
\be
\ydot = c_0 f(y(\t)) \; .
\lab{y-eq-cosmolog}
\ee
Eq.\rf{X0-eq-1} reduces in the case of \rf{cosmolog-type} to:
\be
\pa_\t \chi + \chi\,\frac{\pa_\t S}{S}\,(1-2p\,a_0) = 0 \quad \to \quad
\chi (\t) = \chi_0 \bigl( S(\t)\bigr)^{2p\, a_0-1} \; .
\lab{chi-eq-cosmolog}
\ee
Here again, for the special choice of the integration constant $M$ \rf{phi-eqs}
such that the constant $a_0$ \rf{a0-const} is fine-tuned as $a_0 = \frac{1}{2p}$, 
we obtain a constant solution $\chi = \chi_0$.

\section{Examples}

As a first example of lightlike brane tension's ``inflation''/``deflation''
\rf{chi-eq-kruskal-sol}
let us consider de Sitter embedding space metric in Kruskal-like
(Gibbons-Hawking) coordinates \ct{gibbons-hawking}:
\br
ds^2 = A(y^2-t^2)\llb - (dt)^2 + (dy)^2\rrb +
R^2 (y^2-t^2) g_{ij}(\vec{\th}) d\th^i d\th^j \; ,
\lab{gibbons-hawking-metric} \\
A(y^2-t^2) = \frac{4}{K(1+y^2-t^2)^2} \quad ,\quad
R (y^2-t^2) = \frac{1}{\sqrt{K}}\,\frac{1-(y^2-t^2)}{1+y^2-t^2} \; .
\lab{gibbons-hawking-metric-1}
\er
Substituting:
\be
A(0)=\frac{4}{K}\;\; ,\;\; C(0) \equiv R^2(0) =\frac{1}{K}\;\; ,\;\; 
C^\pr(0)\equiv 2 R(0) R^{\pr}(0)=-\frac{4}{K}
\lab{A-C-gibbons-hawking}
\ee
into expression \rf{chi-eq-kruskal-sol} we get for the dynamical brane tension 
(recall that the cosmological constant $K$ from \rf{gibbons-hawking-metric-1} 
and the constant $a_0$ \rf{a0-const} are strictly positive):
\be
\chi(\t) = \chi_0 \exp\bigl\{\t^2\,p\,a_0 K\bigr\} \; ,
\lab{chi-gibbons-hawking}
\ee
\textsl{i.e.}, exponential ``inflation'' at $\t \to \pm \infty$ for the brane 
tension of lightlike branes occupying de Sitter horizon.

The second example is Schwarzschild background metric in Kruskal coordinates
\ct{kruskal-1}-\ct{kruskal-2},\ct{MTW} (here we take $D=p+2=4$, \textsl{i.e.}, 
$i,j=1,2$):
\br
ds^2 = A(y^2-t^2)\llb - (dt)^2 + (dy)^2\rrb +
R^2 (y^2-t^2) g_{ij}(\vec{\th}) d\th^i d\th^j \; ,
\lab{kruskal-metric}\\
A = \frac{4R_0^3}{R} \exp\Bigl\{ - \frac{R}{R_0}\Bigr\} \quad ,\quad
\Bigl(\frac{R}{R_0}-1\Bigr) \exp\Bigl\{\frac{R}{R_0}\Bigr\} = y^2 - t^2
\quad ,\quad R_0 \equiv 2G_N m
\lab{kruskal-metric-1}
\er
Calculating $A(0)\,,\, C(0)\equiv R^2(0)$ and $C^\pr(0)\equiv 2 R(0) R^{\pr}(0)$
from \rf{kruskal-metric-1} we obtain for \rf{chi-eq-kruskal-sol}:
\be
\chi(\t) = \chi_0 \exp\Bigl\{-\t^2\,\frac{a_0}{R^2_0}\Bigr\} \; ,
\lab{chi-kruskal}
\ee
\textsl{i.e.}, exponential ``deflation'' at $\t \to \pm \infty$ for the brane 
tension of lightlike branes sitting on the Schwarzschild horizon.

Next, we consider Reissner-Nordstr{\"om} background metric in two different
Kruskal-like coordinate systems of the general form 
(here again we take $D=p+2=4$, \textsl{i.e.}, $i,j=1,2$):
\be
ds^2 = A (y^2-t^2)\llb - (dt)^2 + (dy)^2\rrb +
R^2 (y^2-t^2) g_{ij}(\vec{\th}) d\th^i d\th^j \; .
\lab{RN-metric}
\ee
The first one is appropriate for the region around the outer Reissner-Nordstr{\"om}
horizon $R=R_{(+)}$, \textsl{i.e.}, for $R>R_{(-)}$, the latter being the inner
$R=R_{(-)}$ Reissner-Nordstr{\"om} horizon:
\br
y^2 -t^2 = \frac{R-R_{(+)}}{(R-R_{(-)})^{R^2_{(-)}/R^2_{(+)}}}\,
\exp\Bigl\{ R \frac{R_{(+)}-R_{(-)}}{R^2_{(+)}} \Bigr\} \; ,
\lab{RN-kruskal-outer-1} \\
A (y^2-t^2) = 
\frac{4 R^4_{(+)} (R-R_{(-)})^{1+R^2_{(-)}/R^2_{(+)}}}{(R_{(+)}-R_{(-)})^2 R^2}\,
\exp\Bigl\{ - R \frac{R_{(+)}-R_{(-)}}{R^2_{(+)}} \Bigr\} \; .
\lab{RN-kruskal-outer-2}
\er
Accordingly, the second Kruskal-like coordinate system is appropriate for the 
region around the inner Reissner-Nordstr{\"om} horizon $R=R_{(-)}$, 
\textsl{i.e.}, for $R<R_{(+)}$:
\br
y^2 -t^2 = \frac{R-R_{(-)}}{(R-R_{(+)})^{R^2_{(+)}/R^2_{(-)}}}\,
\exp\Bigl\{ R \frac{R_{(-)}-R_{(+)}}{R^2_{(-)}} \Bigr\} \; ,
\lab{RN-kruskal-inner-1} \\
A (y^2-t^2) = 
\frac{4 R^4_{(-)} (R_{(+)}-R)^{1+R^2_{(+)}/R^2_{(-)}}}{(R_{(-)}-R_{(+)})^2 R^2}\,
\exp\Bigl\{ - R \frac{R_{(-)}-R_{(+)}}{R^2_{(-)}} \Bigr\} \; .
\lab{RN-kruskal-inner-2}
\er

Formula \rf{chi-eq-kruskal-sol} for the brane tension in the case of
\rf{RN-kruskal-outer-1}--\rf{RN-kruskal-outer-2} specializes to:
\be
\chi(\t) = \chi_0 \exp\Bigl\{-\t^2\,\frac{a_0}{R^2_{(+)}}
\Bigl( 1 - \frac{R_{(-)}}{R_{(+)}} \Bigr) \Bigr\} \; ,
\lab{chi-RN-outer-1}
\ee
\textsl{i.e.}, we find exponentially ``deflating'' tension for a lightlike brane
sitting on the outer Reissner-Nordstr{\"om} horizon -- a phenomenon similar
to the case of lightlike brane sitting on Schwarzschild horizon \rf{chi-kruskal}.
In the case of \rf{RN-kruskal-inner-1}--\rf{RN-kruskal-inner-2} formula
\rf{chi-eq-kruskal-sol} becomes:
\be
\chi(\t) = \chi_0 \exp\Bigl\{\t^2\,\frac{a_0}{R^2_{(-)}}
\Bigl( \frac{R_{(+)}}{R_{(-)}} - 1 \Bigr) \Bigr\} \; ,
\lab{chi-RN-outer-2}
\ee
\textsl{i.e.}, we obtain exponentially ``inflating'' tension for a lightlike brane
occupying the inner Reissner-Nordstr{\"om} horizon -- an effect similar to
the exponential brane tension ``inflation'' on de Sitter horizon 
\rf{chi-gibbons-hawking}. In the case of extremal Reissner-Nordstr{\"om} horizon, 
\textsl{i.e.} when $R_{(+)}=R_{(-)}$, where both ``deflating'' \rf{chi-RN-outer-1} 
and ``inflating'' \rf{chi-RN-outer-2} solutions should match,
the only solution for the brane tension is the constant one $\chi = \chi_0$.  

Finally, as an example for ``inflation''/``deflation'' behavior of the dynamical 
lightlike brane tension $\chi$ in cosmological-type embedding space-time
\rf{FRW-metric} let us consider Friedman-Robertson-Walker metrics, \textsl{i.e.}, 
background metrics of the form \rf{FRW-metric}, where (see \textsl{e.g.} 
\ct{hawking-ellis}):
\be
f(y)=y \quad ,\quad f(y)=\sin y \quad ,\quad f(y)=\sinh y \; .
\lab{FRW-f}
\ee
Solving Eqs.\rf{horizon-cosmolog}--\rf{y-eq-cosmolog} yields for each choice 
\rf{FRW-f} of $f(y)$ correspondingly:
\br
f(y)=y \;\; \to \;\; y(\t)=y_0 e^{c_0 \t} 
\;\; ,\;\; S(t) = \pm \frac{1}{c_0\, y_0}\, e^{-c_0\, t} \; ,
\lab{FRW-1}\\
f(y)=\sin y \;\; \to \;\; y(\t)= 2\arctan\( e^{c_0 (\t +\t_0)}\)
\;\; ,\;\; 
S(t) = \pm \frac{1}{c_0}\,\cosh\(c_0 (t+\t_0)\)  \; ,
\lab{FRW-2}\\
f(y)=\sinh y \;\; \to \;\; y(\t)= 
\ln\frac{1 + e^{-c_0 (\t +\t_0)}}{1 - e^{-c_0 (\t +\t_0)}} \;\; ,\; c_0 >0
\;\; ,\;\; S(t) = \mp \frac{1}{c_0}\,\sinh\(c_0 (t+\t_0)\) \; ,
\lab{FRW-3}
\er
where $y_0,\t_0 = \mathrm{const}$.
Inserting the expressions \rf{FRW-1}--\rf{FRW-3} for $S(t)$ into 
Eq.\rf{chi-eq-cosmolog} yields a time-asymmetric ``inflation''/``deflation''
of the brane tension $\chi$ at $\t \to \pm\infty$, except for the ``fine tuned''
case $a_0 =\frac{1}{2p}$ where we get a constant $\chi = \chi_0$. 

Let us recall that the metrics \rf{FRW-metric} with any of the three choices 
\rf{FRW-f} for $f(y)$ and the corresponding expressions for $S(t)$ given by
\rf{FRW-1}--\rf{FRW-3} represents de Sitter space-time in various
coordinatizations different from the Gibbons-Hawking one
\rf{gibbons-hawking-metric}--\rf{gibbons-hawking-metric-1} 
(here $|c_0| = K$ with $c_0$ and $K$ from \rf{FRW-1}--\rf{FRW-3}
and \rf{gibbons-hawking-metric-1}, respectively).
Let us also stress the qualitative difference between the solutions for the
brane tension of lightlike branes occupying de Sitter horizons:
time-asymmetric ``inflation''/``deflation'' behavior \rf{chi-eq-cosmolog}
with  exponential linear time dependence in Friedman-Robertson-Walker
coordinates versus strictly ``inflationary'' behavior \rf{chi-gibbons-hawking}
with exponential quadratic time dependence in Gibbons-Hawking (Kruskal-like)
coordinates.

\section{Discussion and Conclusions}

In the present paper we presented a systematic Lagrangian formulation of
lightlike $p$-branes in arbitrary $(p+1)$ world-volume dimensions, whose
brane tension becomes an additional nontrivial dynamical degree of freedom.
Further, we have shown that codimension one lightlike branes can move 
in gravitational backgrounds of spherically symmetric type provided the latter
possess
event horizons and, moreover, these horizons are automatically occupied 
(``straddled'') by the lightlike branes.


For more conventional type of matter, a process known as ``mass inflation''
\ct{poisson-israel-1}-\ct{poisson-israel-2} leads to matter accumulation on 
certain horizons (like the
inner Reissner-Nordstr{\"o}m horizon) and, therefore, is somewhat similar to
the phenomenon of automatic positioning of lightlike branes on black hole or
cosmological horizons. For the standard ``mass inflation'' one defines a
mass function (not related to the external mass of the black hole) which
grows without bound as the matter focuses on the horizon.
The natural analog of the mass function in the case of lightlike branes
appears to be the dynamical brane surface tension. We study the time
dependence of the dynamical brane tension of lightlike branes occupying
diverse horizons.

Employing appropriate ans{\"a}tze for various sets of Kruskal-like coordinates
(Gibbons-Hawking coordinates \ct{gibbons-hawking} in the case of de Sitter space)
we find solutions for the lightlike branes of \rf{LL-brane} located at the inner 
Reissner-Nordstr{\"o}m 
horizon or at the de Sitter cosmological horizon, respectively, such that the
dynamical brane tension undergoes time-reflection symmetric ``mass inflation'',
\textsl{i.e.}, it approaches exponentially arbitrary large values for
$\t \to \pm\infty$. Although the present result for dynamical brane tension
``inflation'' at the inner 
Reissner-Nordstr{\"o}m horizon parallels (except for the time-reflection
symmetry here obtained) the known ``mass inflation'' phenomenon for standard
matter, the accompanying result about brane tension ``inflation'' at de Sitter space
horizon represents something totally new with no analog within the standard matter 
``mass inflation'' and, therefore, it is a unique feature of lightlike branes.

In contrast, using the same ans{\"a}tze with Kruskal-like coordinates, we find 
that lightlike branes undergo ``mass deflation'', \textsl{i.e.}, their dynamical 
brane tension going to zero for $\t \to \pm\infty$ when they are located at the
outer Reissner-Nordstr{\"o}m or the Schwarzschild horizon.

Other types of ans{\"a}tze natural for standard coordinates show that for
all kinds of horizons there are time-asymmetric ``mass inflation'' or ``mass
deflationary'' solutions for the dynamical lightlike brane tension and, 
for a fine tuning -- also solutions with constant brane tension do exist.
In particular, for de Sitter horizon in cosmological (Friedman-Robertson-Walker)
coordinates we obtain time-asymmetric ``inflation''/``deflation'' with
exponential linear time dependence in contrast to the strict ``mass inflation''
at Sitter horizon in Gibbons-Hawking (Kruskal-like) coordinates with
exponential quadratic time dependence.

Let us stress that in the present paper we have discussed the properties of 
\textsl{LL-brane} dynamics as {\em test branes} moving in various gravitational 
backgrounds, \textsl{i.e.}, we have not taken into account the back-reaction
of \textsl{LL-branes} on the geometry and the physical properties of the embedding
space-time. In a forthcoming paper we are studying the important issue of 
self-consistent solutions for bulk gravity-matter systems (\textsl{e.g.}, 
Einstein-Maxwell-type)
coupled to lightlike branes, \textsl{i.e.}, accounting for its back-reaction,
where the latter: (i) serve as a source for gravity
and electromagnetism, (ii) dynamically produce space-varying cosmological 
constant, and (iii) trigger non-trivial matching of two different geometries of 
de-Sitter/black-hole type across common horizon spanned by the lightlike
brane itself. In fact, we have already started the above study in our
previous papers \ct{will-brane-1},\ct{will-brane-5}-\ct{will-brane-7} in the
simplest case of horizon matching of two different spherically-symmetric
space-times where the pertinent lightlike brane occupying the common horizon
has {\em constant} dynamical tension (``static soldering'' in the terminology of
ref.\ct{Barrabes-Israel-Hooft-1}). One of the physically interesting cases is a solution with de Sitter interior
region with dynamically generated cosmological constant through the coupling
to the \textsl{LL-brane}, and an outer region with Schwarzschild or
Reissner-Nordstr{\"o}m geometry, \textsl{i.e.}, a {\em non-singular} black
hole. 

\section*{Acknowledgments}

E.N. and S.P. are supported by European RTN network
{\em ``Forces-Universe''} (contract No.\textsl{MRTN-CT-2004-005104}).
They also received partial support from Bulgarian NSF grants \textsl{F-1412/04}
and \textsl{DO 02-257}.
Finally, all of us acknowledge support of our collaboration through the exchange
agreement between the Ben-Gurion University of the Negev (Beer-Sheva, Israel) and
the Bulgarian Academy of Sciences.


\end{document}